\documentclass{PoS}
\usepackage{lineno}

\title{Searches for LFV with CMS: leptoquarks with couplings to quarks of the 3rd generation}

\ShortTitle{Searches for leptoquarks with couplings to quarks of the 3rd generation in CMS}

\author{\speaker{Johannes Haller} on behalf of the CMS collaboration\\
  Institut f\"ur Experimentalphysik, Universit\"at Hamburg, Germany\\
  E-mail: \email{Johannes.Haller@physik.uni-hamburg.de}}

\abstract{Leptoquarks (LQ) with couplings to the third generation of
  Standard Model (SM) quarks have been proposed as possible
  explanations of the flavour anomalies indicating the violation of
  lepton flavour universality (LFV). The CMS collaboration has
  initiated an extensive search programme for these new states in the
  LHC run-2 data recorded at $\sqrt{s}=13$\,TeV. In this article, the
  CMS search results in the LQ pair-production final states $\tau
  b\,\tau b$, $\nu t\,\nu t$, $\nu b\,\nu b$, $\tau t\,\tau t$, and
  $\mu t\,\mu t$, as well as in the LQ single-production final state
  $\tau\,\tau b$ are discussed. No significant deviation from the SM
  is observed in any of these channels. For a broad range of LQ decay
  modes, exclusion limits on the LQ masses are determined at 95\%
  confidence level reaching from 1.0\,TeV to 1.8\,TeV.}

\FullConference{Sixth Annual Conference on Large Hadron Collider Physics (LHCP2018)\\
                4-9 June 2018\\
                Bologna, Italy}

\begin{document}

\section{Introduction}

While direct searches for new physics at the LHC have not seen any
significant deviation from the Standard Model (SM) prediction so far,
anomalies in the flavour sector have been reported pointing towards a
violation of lepton flavour universality (see
e.g.~\cite{crivellin,mogini,Amhis:2016xyh} and references therein). In
particular, the measurements of rare decays of $B$ mesons by the
BABAR, BELLE and LHCb collaborations feature several interesting
observations. Among these, the ratios $R_D$ and $R_{D^*}$, defined as
ratios of the branching fractions for $\overline{B} \rightarrow
D^{(*)}\tau\overline{\nu}$ to those for $\overline{B} \rightarrow
D^{(*)}\mu\overline{\nu}$ show a deviation with a combined significance of about
four standard deviations from the SM prediction
(see~\cite{Amhis:2016xyh} and references therein). The ratio $R_K$ of
the branching fractions for $B\rightarrow K^{(*)}\mu\mu$ to those for
$B\rightarrow K^{(*)}ee$ shows a deviation of 2.6 standard
deviations~\cite{Aaij:2014ora}.

As a possible explanation for these deviations, leptoquarks (LQ) with
couplings to SM quarks of the third generation ($b,t$) have been
proposed (see e.g.~\cite{crivellin} and references therein). LQs are
hypothetical new scalar (spin $J=0$) or vector ($J=1$) particles which
decay to a lepton and a quark and carry strong and fractional
electromagnetic charges. They appear in many extensions of the SM,
like GUT-inspired models, technicolour, compositeness or $R$-parity
violating supersymmetry. Classification of different LQ types is
usually performed within the Buchm\"uller-R\"uckl-Wyler
model~\cite{Buchmuller:1986zs} depending on their quantum
numbers. Often the existence of only three categories of LQs is
assumed, each featuring couplings to leptons and quarks of the same SM
generation. However, relaxing these coupling restrictions within the
constraints on flavour changing neutral currents and other rare
processes, would lead to final states with leptons and quarks from
different SM generations.

At the LHC, processes of the strong interaction (gluon-gluon fusion or
quark-antiquark annihilation) lead to a copious pair-production of LQs
resulting in $l^{\pm}q\,l^{\mp}q$, $l^{\pm}q\,\nu q'$ and $\nu q'\,\nu
q'$ final states depending on the LQ branching fraction to a quark and
a charged lepton, $\beta$, or to a quark and a neutrino,
$1-\beta$. The cross-sections of the strong pair-production only
depend on the LQ mass $M_{\rm LQ}$ and are independent of the
Yukawa coupling $\lambda$ of the LQ-lepton-quark vertex. Calculations
at leading order (LO) and next-to-leading order (NLO) accuracy are available for this production
mode for vector and scalar leptoquarks~\cite{Dorsner:2018ynv,Kramer:2004df}, respectively.  In general, the single-production of LQs
via quark-gluon scattering is suppressed at the LHC for LQs with
couplings to third generation quarks, since it requires a heavy quark
in the initial state. However, for high values of $\lambda$ and
$M_{\rm LQ}$ the single-production could be sizeable for LQs with
couplings to bottom quarks leading to $l^{\pm}l^{\pm} b$, $l^{\pm}\nu
b$ and $\nu\nu b$ final states. The cross-section for LQ
single-production is dependent on both $M_{\rm LQ}$ and $\lambda$.

The CMS collaboration has initiated a broad range of direct searches
for the pair-production of LQs with couplings to quarks of the third
generation. CMS results based on the LHC run-2 dataset recorded at a
centre-of-mass energy of $\sqrt{s}=13$\,TeV are available in the $\tau
b\,\tau b$~\cite{Khachatryan:2016jqo,Sirunyan:2017yrk,CMS:2018eud},
$\nu t\,\nu t$~\cite{Sirunyan:2017kqq,Sirunyan:2018kzh}, $\nu b\,\nu
b$~\cite{Sirunyan:2017kqq,Sirunyan:2018kzh}, $\tau t\,\tau
t$~\cite{Sirunyan:2018nkj} and $\mu t\,\mu t$~\cite{CMS:2018itt} final
states.  In addition, a first search for the single-production of LQs
has been performed in the $\tau\,\tau b$ final
state~\cite{Sirunyan:2018jdk}. Results of searches for third
generation LQs from LHC run-2 data by the ATLAS collaboration are not
yet available.

\section{CMS searches for leptoquarks with couplings to 3rd generation quarks}

In the following subsections a short introduction to the CMS results
of direct searches for LQs with couplings to 3rd generation quarks is
given. For more details on the physics background, on the data
analysis, on the results, on the physics interpretation, as well as
for a more complete list of references, the reader is referred to the
original CMS publications.  Unless stated otherwise, all results
discussed in this review are based on a dataset of LHC $pp$ collisions
at $\sqrt{s}=13\,$\,TeV recorded by the CMS
experiment~\cite{Chatrchyan:2008aa} in 2016, corresponding to an
integrated luminosity of $35.9\,\rm{fb}^{-1}$.

\subsection{LQ pair-production in $\mathbf{\nu j\,\nu j}$, $\mathbf{\nu b\,\nu b}$ and $\mathbf{\nu t\,\nu t}$ final states}
\label{sec:SUSYreint}
\begin{figure}[t]
     \includegraphics[width=.51\textwidth]{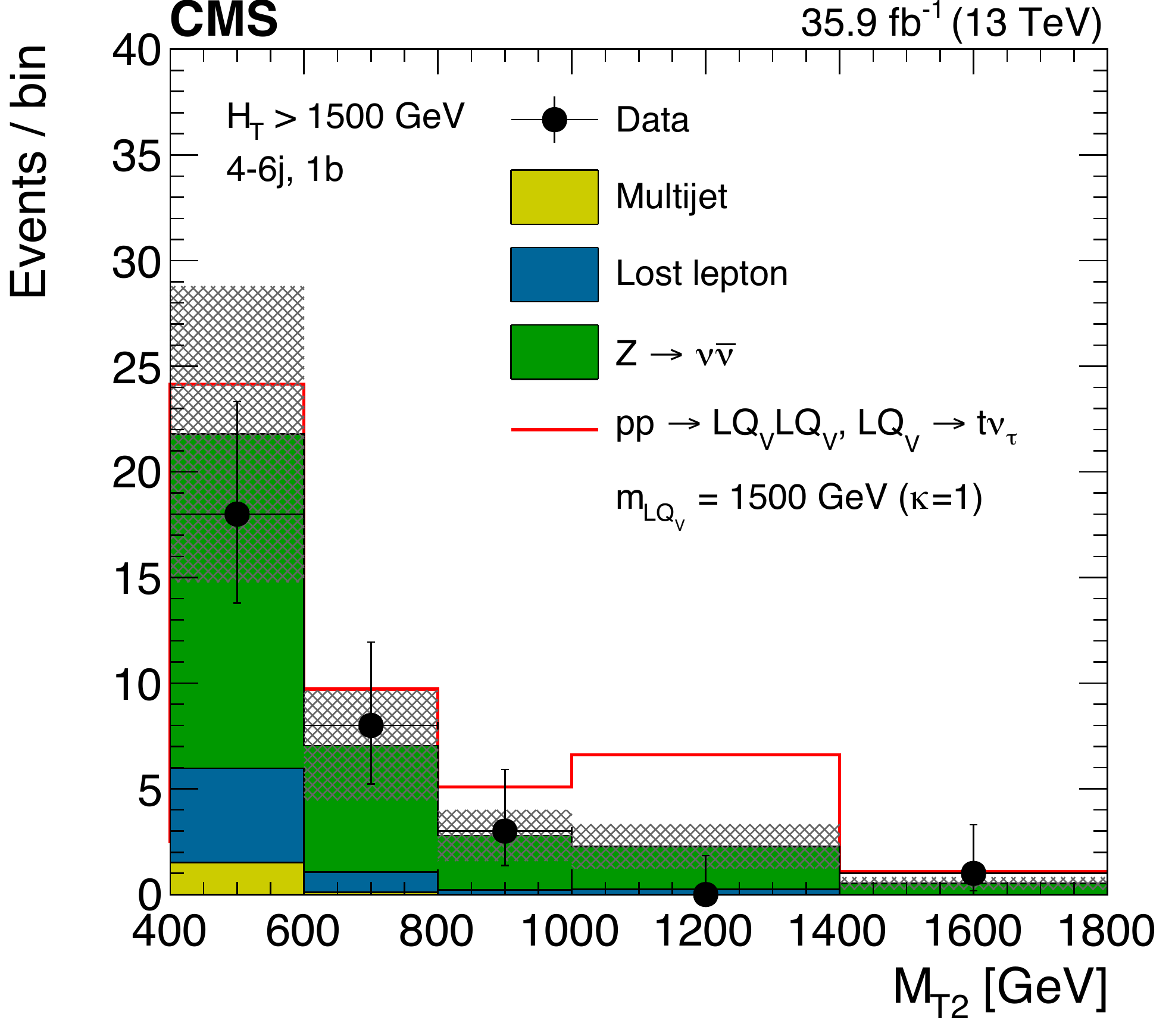}
     \includegraphics[width=.49\textwidth]{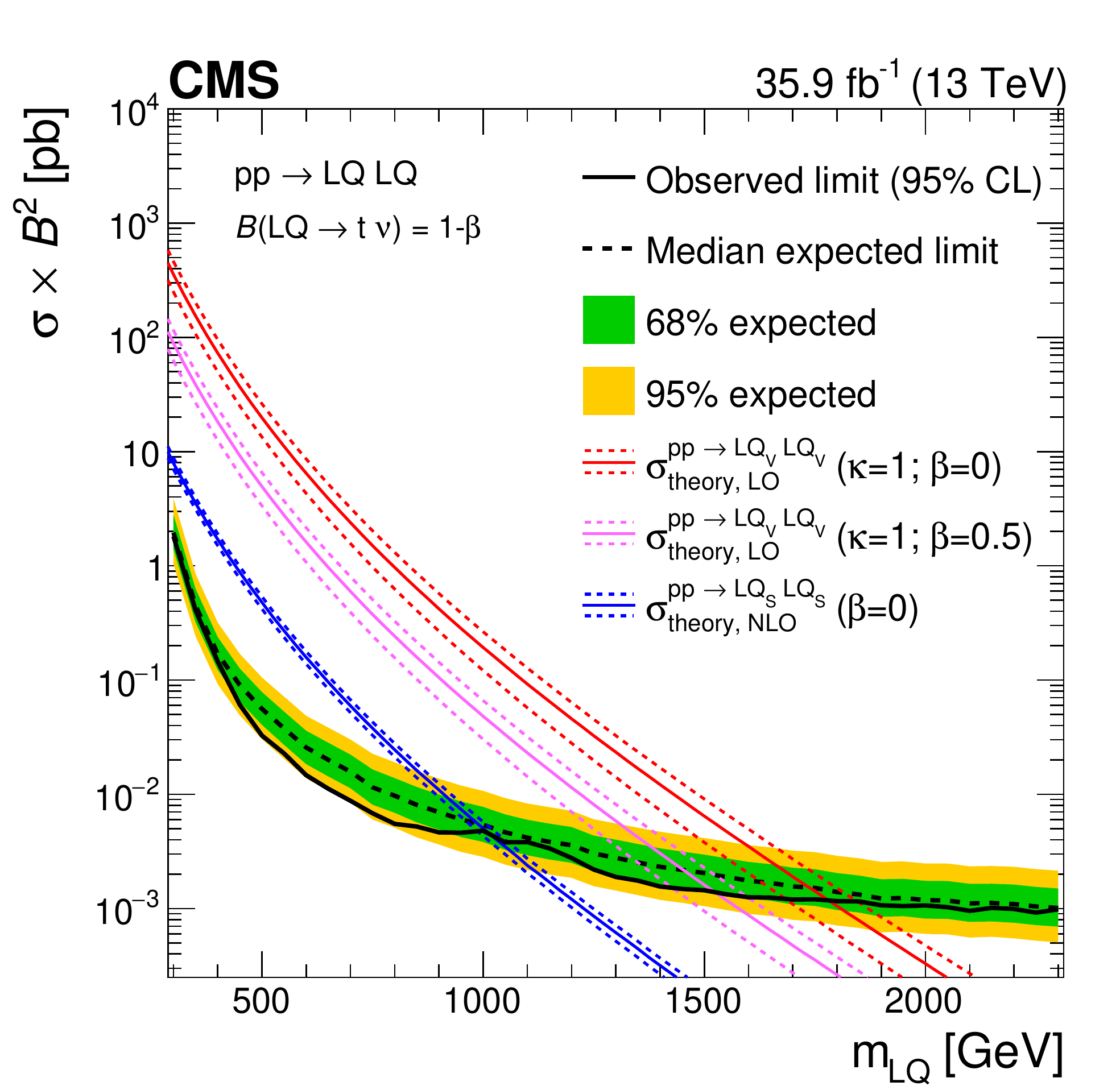}
     \caption{Reinterpretation of a CMS search for squarks
       and gluinos to constrain LQ
       modes~\cite{Sirunyan:2017kqq,Sirunyan:2018kzh}: (left) observed
       distribution of the $M_{\rm T_2}$ variable compared to the
       expectation from SM background processes and a hypothetical LQ
       signal (stacked on top of the background expectation) assuming
       a mass of $M_{\rm LQ}=1500$\,GeV; (right) exclusion limits at
       95\% CL on the production cross section as a function of the
       LQ mass for pair-production of LQs decaying with unit
       branching fraction to a neutrino and a top quark.}
     \label{fig:SUSYreint}
\end{figure}

For the study of LQ pair-production in $\nu j\,\nu j$, $\nu b\,\nu b$,
and $\nu t\,\nu t$ final states, the CMS collaboration uses the
results of an analysis~\cite{Sirunyan:2017kqq} searching for
the SUSY pair-production of squarks in the decay channel
$\tilde{q}\tilde{q}\rightarrow \chi_1^0 q\,\chi_1^0 q$ using events
with jets and significant transverse momentum imbalance, as inferred
through the $M_{\rm T2}$ variable. The analysis makes use of 213
exclusive search regions to cover a broad range of new physics scenarios and
masses. Recently, the search results have been
reinterpreted~\cite{Sirunyan:2018kzh} for the study of LQ
pair-production scenarios (${\rm LQ\,\overline{LQ}}\rightarrow \nu
q\,\nu q$). In Fig.~\ref{fig:SUSYreint}\,(left) the distribution of
the $M_{\rm T_2}$ variable is shown in a signal region with particular
strong sensitivity for the LQ signal. The analysis does not show any
significant deviation from the SM expectation which has been estimated
from dedicated data control regions. Exclusion limits at 95\%
confidence level (CL) on the LQ pair-production cross-section have
been determined for LQ decays to all quark flavours. The kinematic
differences of the pair-production between scalar and vector LQs have
been found to be within the assumed uncertainties. The exclusion
limits on the production cross-section as observed and expected for
the LQ decay into $\nu t$ are shown in
Fig.~\ref{fig:SUSYreint}\,(right) compared to three signal models of
scalar and vector LQs. For $\beta=0$, i.e. for LQs decaying
exclusively into the $\nu q$ channel, scalar (vector) LQs are excluded
up to $M_{\rm LQ}=980\,/ 1100\,/1020$\, GeV (1790\,/ 1810\,/1780\,
GeV) for $q=udsc\,/ b\,/ t$ respectively.

\subsection{LQ pair-production in $\mathbf{\tau t \tau t}$ final
  states}

\begin{figure}[t]
     \includegraphics[width=.435\textwidth]{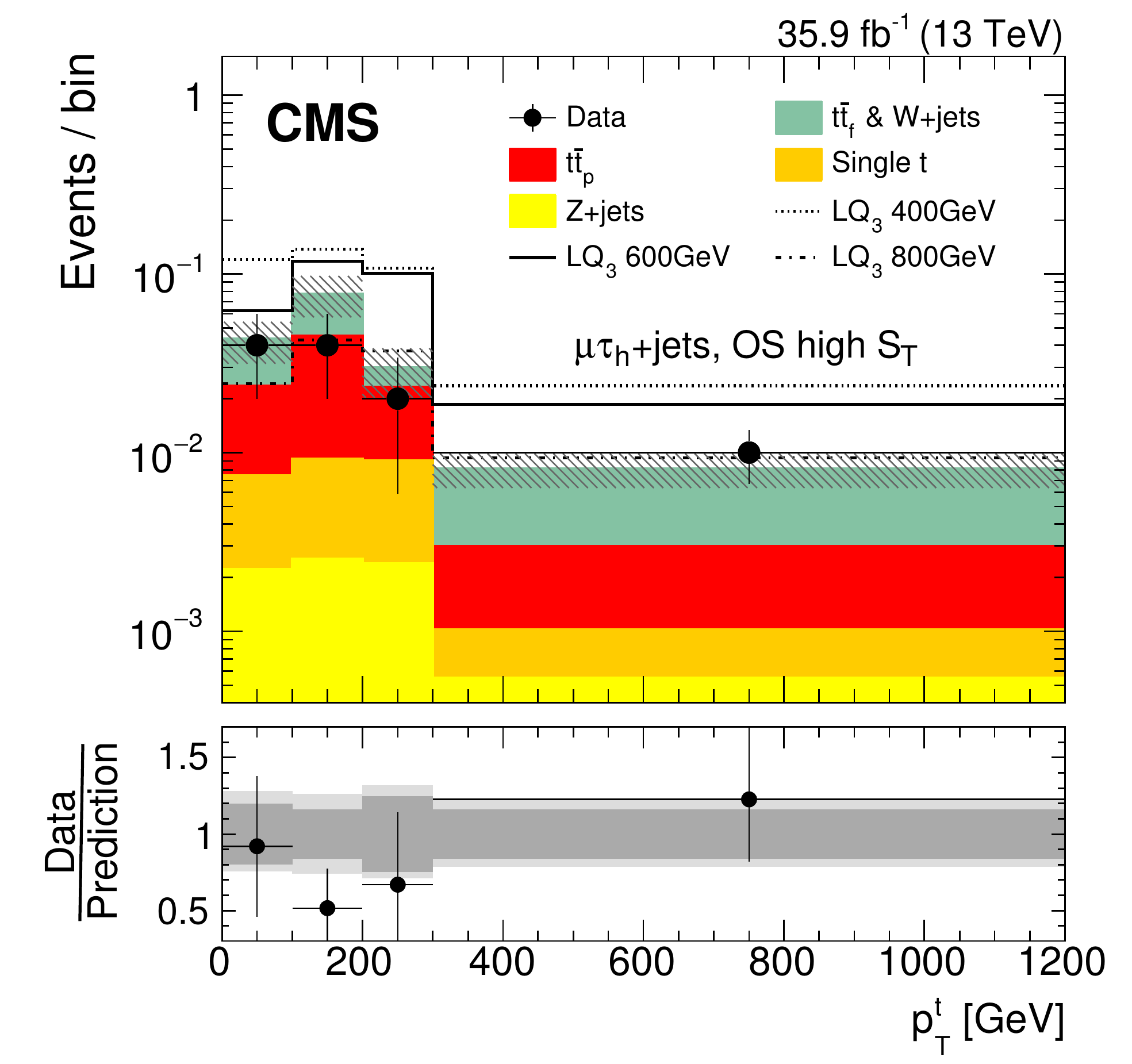}
     \includegraphics[width=.565\textwidth]{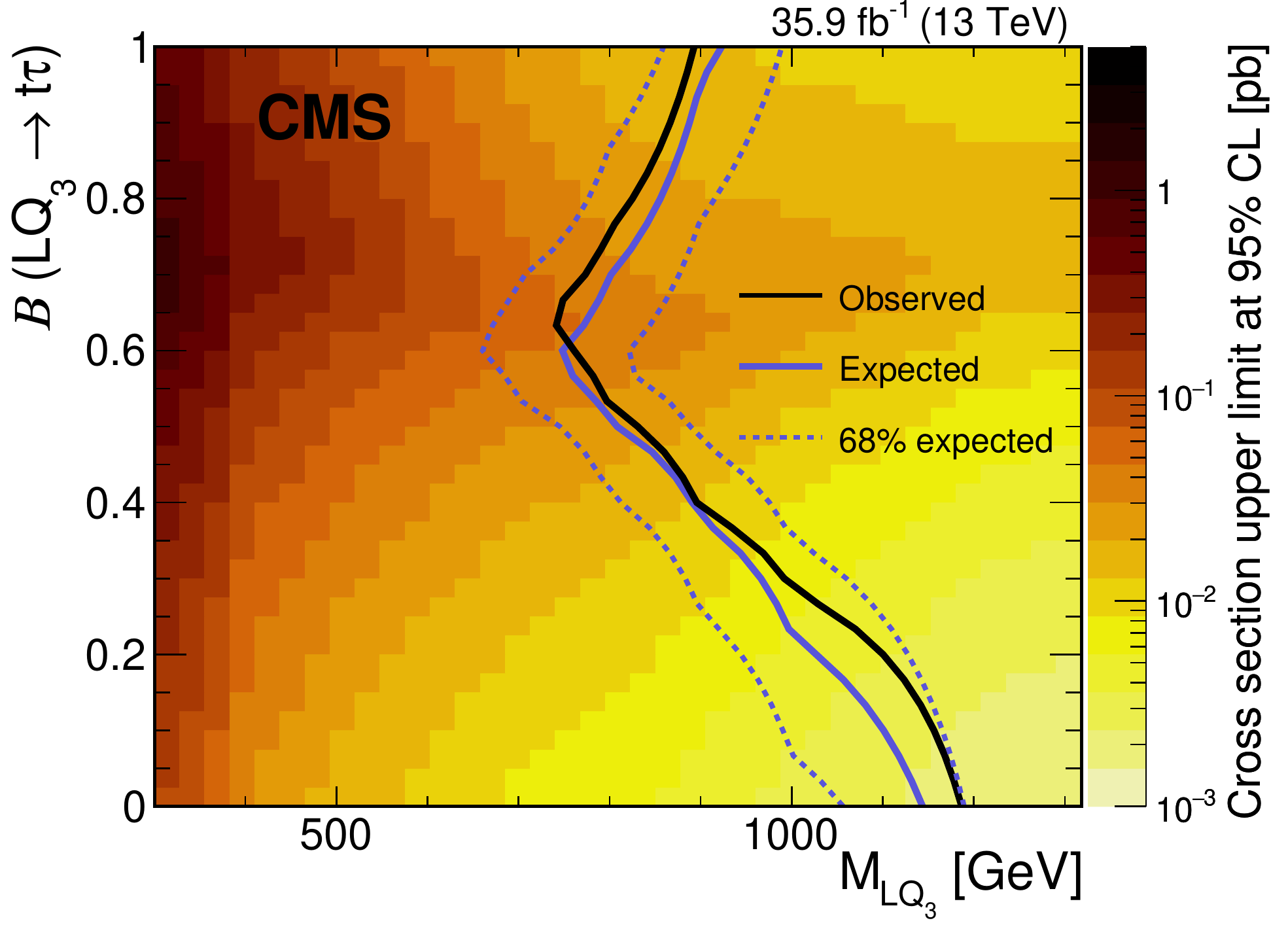}
     \caption{Search for pair-produced LQs with decays to a top quark
       and a $\tau$ lepton~\cite{Sirunyan:2018nkj}: (left)
       distribution of the transverse momentum of the top quark
       candidate in one of the 10 categories considered in the
       analysis; (right) upper exclusion limits at 95\% CL on the
       production cross section for pair-production of scalar LQs decaying
       into a top quark and a $\tau$ lepton ($B=1$) or a bottom quark
       and a neutrino ($B=0$) in the ($M_{\rm LQ},B$)-plane. The solid
       black (blue) line shows the observed (expected) mass exclusion
       limits. The plot additionally includes results from a search
       for pair-produced bottom squarks~\cite{Sirunyan:2017kqq}.}
     \label{fig:toptau}
\end{figure}

In the CMS search for the pair-production of LQs with decays into top
quarks and $\tau$ leptons in the LHC run-2
dataset~\cite{Sirunyan:2018nkj} events are selected with at least one
electron or muon (denoted as $l$), at least one hadronically decaying
$\tau$ lepton (denoted as $\tau_{\rm h}$), and additional hadronic
jets. The first analysis of this LQ decay channel has been performed
by the CMS collaboration using LHC data recorded at
$\sqrt{s}=8$\,TeV~\cite{Khachatryan:2015bsa} leading to a 95\% CL
exclusion limit on the leptoquark mass of 685\,GeV assuming unit
branching fraction into $\tau t$. For the data recorded at
$\sqrt{s}=13$\,GeV, the analysis had to be changed significantly due
to different background contributions and relative impact of
systematic uncertainties. For a maximum expected significance for a
hypothetical LQ signal, the run-2 analysis splits the event sample
into 10 exclusive categories according to the number of $\tau_{\rm h}$
candidates, $N(\tau_{\rm h})$, the relative $l/\tau_{\rm h}$ charges
(OS, SS) and the value of $S_{\rm T}$, defined as the scalar sum of
the transverse momenta of all selected leptons, jets and missing
transverse momentum. In all categories a considerable fraction of
events is selected in which jets are misidentified as $\tau_{\rm h}$
candidates. For the estimation of this contribution in $t\overline{t}$
and $W$+jets events, a sophisticated data-driven method is employed
depending on the category.

Since only a small number of events is selected with $N(\tau_{\rm
  h})\ge 2$, simple counting experiments are performed for the
corresponding two ($e$ or $\mu$) signal categories. In the electron
channel 9 events are observed in the data, while $7.9^{+2.4}_{-2.5}$
are expected from SM background processes. In the muon channel 11
events are observed in the data, while $8.4^{+2.6}_{-2.3}$ are
expected. In the eight remaining signal categories with $N(\tau_{\rm
  h})=1$, the four-momentum of an hadronically decaying top quark is
reconstructed and the distribution of its transverse momentum
($p^t_{\rm T}$) is used in a statistical analysis of the shape
observed in the data and expected for the SM backgrounds and the LQ
signal. An example $p^t_{\rm T}$ distribution in one of the most
sensitive categories is shown in Fig.~\ref{fig:toptau}\,(left).

No significant deviation from the background expectation is found in
any of the 10 signal categories and exclusion limits at 95\% CL have
been obtained from a binned maximum likelihood (ML) fit. Assuming a
unit braching fraction of the LQs into $\tau t$, masses of scalar
leptoquarks are excluded up to 900\,GeV.  The results of this search
have been combined with a CMS search for the SUSY pair-production of
scalar sbottom quarks in the decay mode
$\tilde{b}\,\tilde{b}\rightarrow \chi_1^0 b \chi_1^0 b$
(\cite{Sirunyan:2017kqq}, see Sect.~\ref{sec:SUSYreint}) which can be
interpreted as a search for LQ pair-production in the decay mode ${\rm
  LQ\,\overline{LQ}} \rightarrow \nu b \nu b$ representing the complementary LQ
decay channel which opens for branching fractions smaller than 1. In
Fig.~\ref{fig:toptau}\,(right) the upper exclusion limits on the
pair-production cross-section for scalar LQs obtained from this
combination is shown in the (mass, branching fraction)-plane. For all
values of the branching fraction, masses of scalar LQs are excluded up
to 800\,GeV.

\subsection{LQ pair-production in $\mathbf{\mu t \mu t}$ final states}

\begin{figure}[t]
     \includegraphics[width=.475\textwidth]{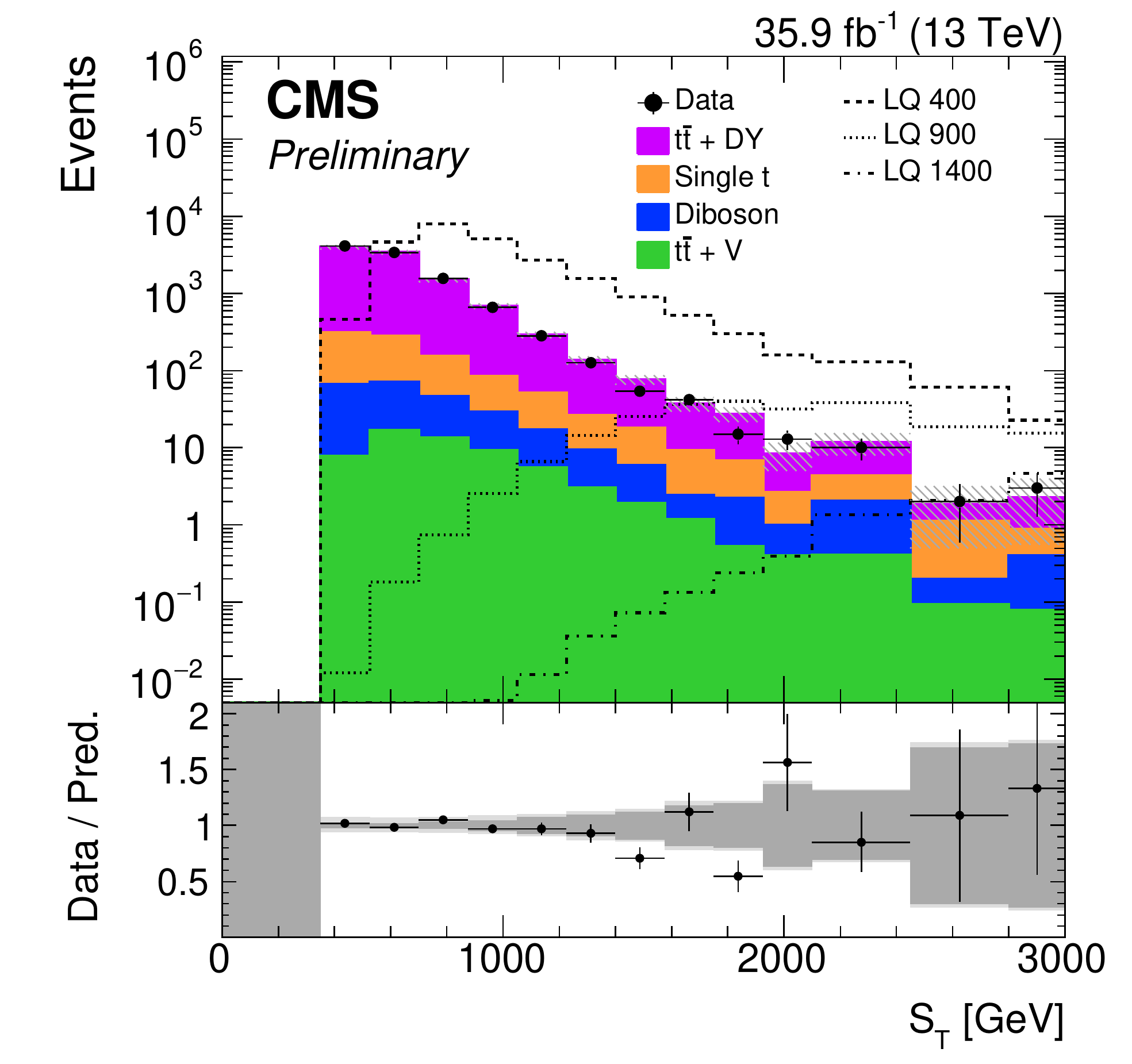}
     \includegraphics[width=.525\textwidth]{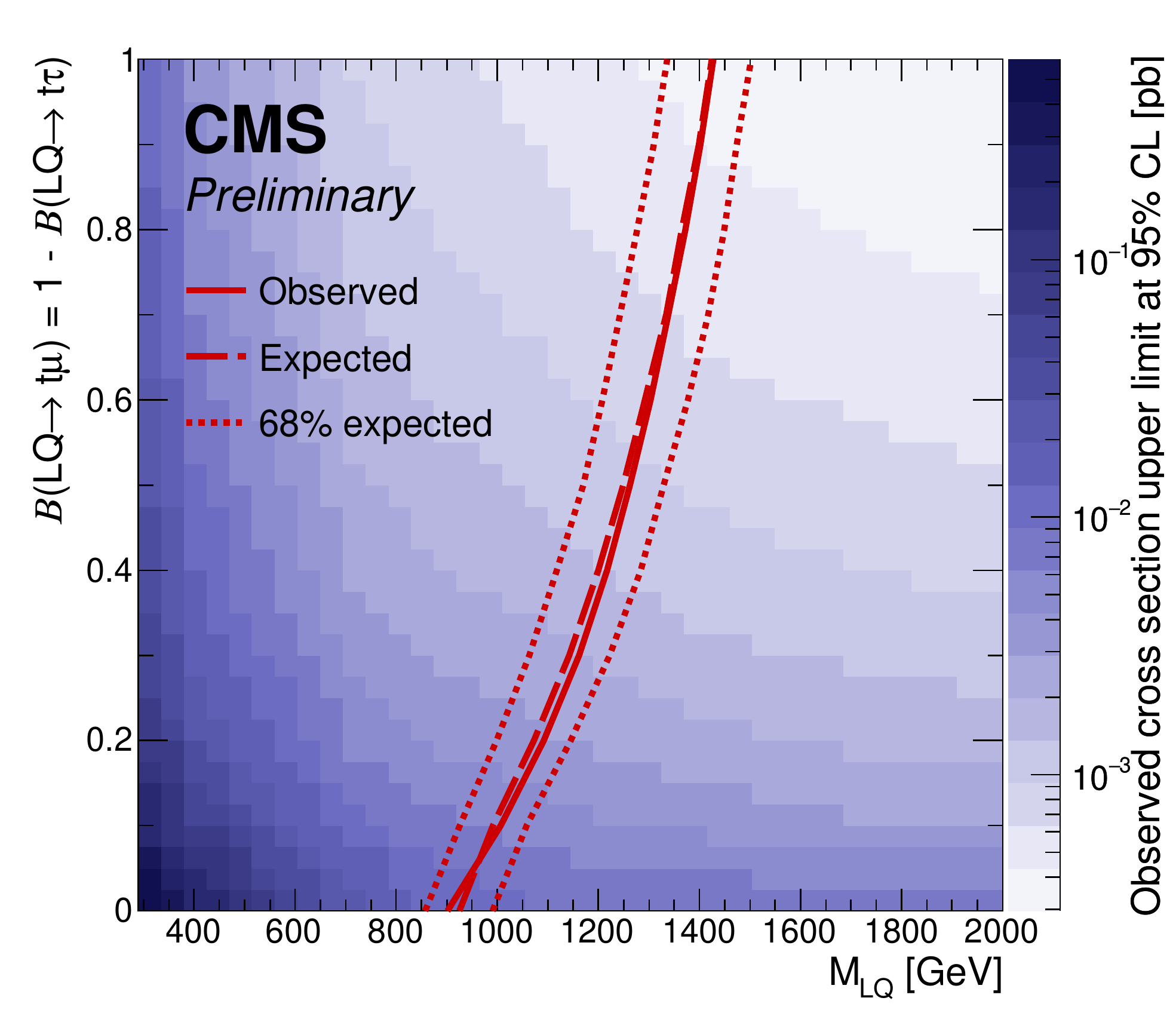}
     \caption{Search for pair-produced LQs with decays to a top quark
       and a muon~\cite{CMS:2018itt}: (left) observed $S_{\rm T}$
       distribution compared to the post-fit expectations from SM
       background processes and hypothetical LQ signals assuming three
       different LQ masses; (right) upper exclusion limits at 95\% CL
       on the production cross section for pair-production of scalar
       LQs decaying into a top quark and a muon ($B=1$) or a top quark
       and a $\tau$ lepton ($B=0$) in the ($M_{\rm LQ},B$)-plane. The
       solid (dashed) line shows the observed (expected) mass
       exclusion limits.  The plot additionally includes results from
       the search for pair-produced LQs with decays to top quarks and
       $\tau$ leptons~\cite{Sirunyan:2018nkj}.}
     \label{fig:topmu}
\end{figure}

The CMS search for LQ pair-production in the decay mode $\mu t\,\mu t$
in the run-2 dataset~\cite{CMS:2018itt} represents the first search
for this LQ type which could provide a simultaneous explanation of
deviations in $R_{D^{(*)}}$, $R_K$, and the anomalous magnetic moment
of the muon (see list of references in \cite{CMS:2018itt}). Events
with at least two muons and at least two hadronic jets are split into
two exclusive signal categories to obtain sensitivity for a broad
range of LQ masses. For events with an additional electron or muon,
the mass of the LQ can be reconstructed using permutations of jets and
lepton candidates which best fit the decay hypothesis for the
considered LQ mode. Since the reconstructed average LQ mass ($M_{\rm
  LQ}^{\rm rec}$) shows strong discrimination power between signal and
SM background, its distribution has been used for the final
statistical interpretation in this category. No significant deviation
from the SM expectation has been observed in this category which
features a particularly strong sensitivity for smaller LQ masses. An
improved significance at higher LQ masses is expected for the
remaining events (i.e. those without additional lepton). In this category,
the spectrum of $S_{\rm T}$, shown in Fig.~\ref{fig:topmu}\,(left),
was chosen for the statistical interpretation. Again, a significant
deviation from the SM expectation has not been observed. In both
categories, the contribution of SM background processes has been estimated using
sophisticated data-driven methods.

Exclusion limits on the LQ pair-production cross-section have been
obtained from simultaneous fits of the $M_{\rm LQ}^{\rm rec}$ and
$S_{\rm T}$ distributions. Assuming unit LQ branching fraction into
$\mu t$, scalar LQs with masses up to 1420\,GeV can be excluded. In
addition, the analysis results have been combined with the analyses of
the $\tau t$~\cite{Sirunyan:2018nkj} and the $\nu
b$~\cite{Sirunyan:2017kqq} decay modes. An example for the resulting
limits on the pair-production cross-section of scalar LQs in the
(mass, branching fraction)-plane is shown in
Fig.~\ref{fig:topmu}\,(right). Scalar LQs with masses up to 900\,GeV
are excluded at the 95\% CL for all values of the branching fractions
into $\mu t$, $\tau t$ and $\nu b$ (not shown).

\subsection{LQ pair-production in $\mathbf{\tau b\,\tau b}$ final states}
\label{sec:taub}
\begin{figure}[t]
     \includegraphics[width=.502\textwidth]{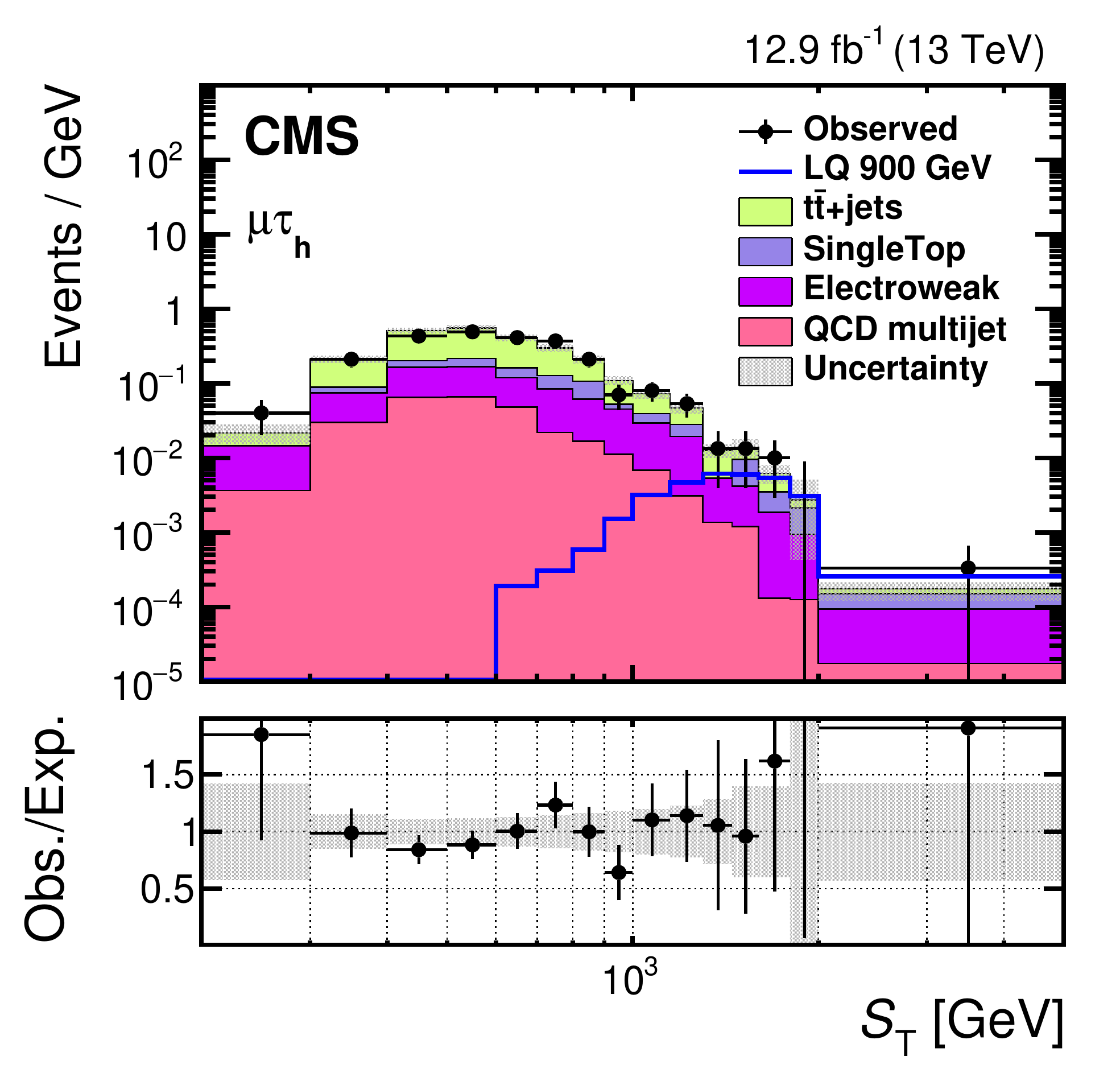}
     \includegraphics[width=.498\textwidth]{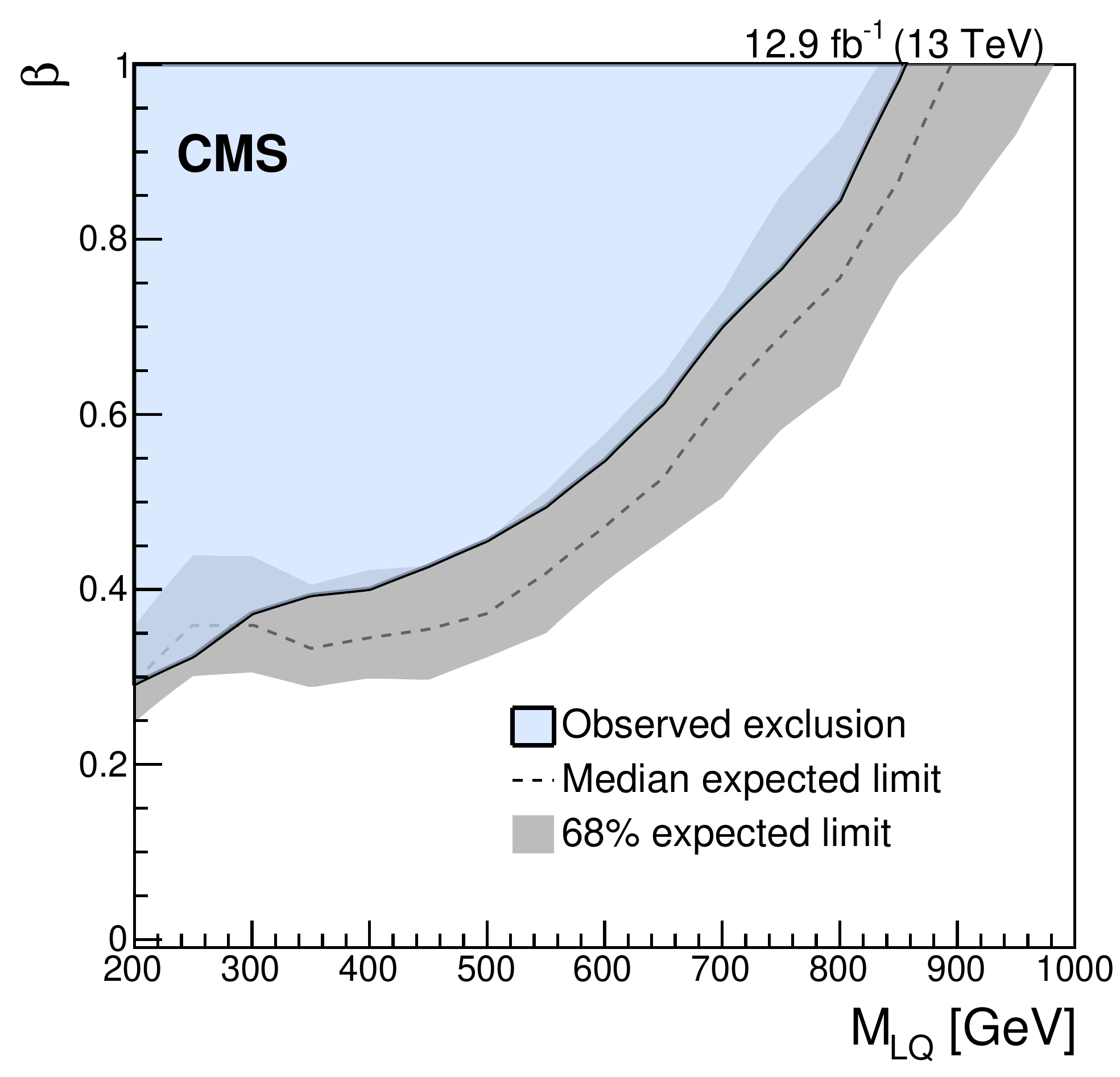}
     \caption{Search for pair-produced LQs with couplings to $\tau$
       leptons and bottom quarks~\cite{Sirunyan:2017yrk}: (left)
       observed $S_{\rm T}$ distribution in the $\mu \tau_h$ channel
       compared to the expectation from SM background processes and a
       hypothetical LQ signal assuming a mass of $M_{\rm
         LQ}=900$\,GeV; (right) exclusion limits at 95\% CL on the
       branching fraction $\beta$ of the LQ to a $\tau$ lepton and a
       bottom quark as a function of the LQ mass.}
     \label{fig:taub}
\end{figure}

For the search for LQ pair-production in the ${\tau b\,\tau b}$ final
state, several results have been published by the CMS collaboration
based on different subsets of the LHC run-2 data at
$\sqrt{s}=13$\,TeV. An analysis~\cite{Khachatryan:2016jqo} of the
dataset recorded in 2015 ($2.1\,{\rm fb}^{-1}$) in the $\tau_{\rm h}
b\,\tau_{\rm h} b$ channel has found no significant deviation from the
SM expectation and has resulted in an initial 95\% CL mass exclusion
limit for scalar LQs of 740\,GeV for unit branching fraction of the LQ
into $\tau b$.

A subset of the full 2016 CMS dataset, corresponding to a reduced
integrated luminosity of only $12.9\,\rm{fb}^{-1}$, has been analysed
in the $\tau_{\rm l} b\,\tau_{\rm h} b$
channel~\cite{Sirunyan:2017yrk}. The analysis relies on a binned ML
fit of the $S_{\rm T}$ distributions in the $e\tau_{\rm h}$ and
$\mu\tau_{\rm h}$ channels. The latter is shown in
Fig.~\ref{fig:taub}\,(left) for illustration. No significant deviation
from the SM expectation is observed and exclusion limits on the LQ
pair-production cross-section are obtained as a function of $M_{\rm
  LQ}$ and $\beta$. A comparison with the theory
cross-section~\cite{Kramer:2004df} provides excluded regions in the
($M_{\rm LQ},\beta$)-plane as shown in
Fig.~\ref{fig:taub}\,(right). The mass limit of 850\,GeV for $\beta=1$
is degraded for lower values of $\beta$ since information in the
complementary $\nu t\,\nu t$ channel has not been included in this
analysis at the time of publication. However, the corresponding mass
limit for $\beta=0$ from the CMS analysis discussed in
Sect.~\ref{sec:SUSYreint} has been determined to be $M_{\rm
  LQ}=1020$\,GeV~\cite{Sirunyan:2018kzh}.

The results in the ${\tau b\,\tau b}$ final state have been updated
during the course of this conference~\cite{CMS:2018eud} using the full
2016 dataset of $35.9\,{\rm fb}^{-1}$. The updated result provides an
observed (expected) exclusion limit on the mass of scalar LQs of
$1.02$ $(1.0)$\,TeV, assuming a 100\,\% branching fraction for the
leptoquark decay into $\tau b$. These results represent the most
stringent limits in this decay channel to date.

\subsection{Single LQ production in $\mathbf{\tau + \tau b}$ final states}
\begin{figure}[t]
     \includegraphics[width=.498\textwidth]{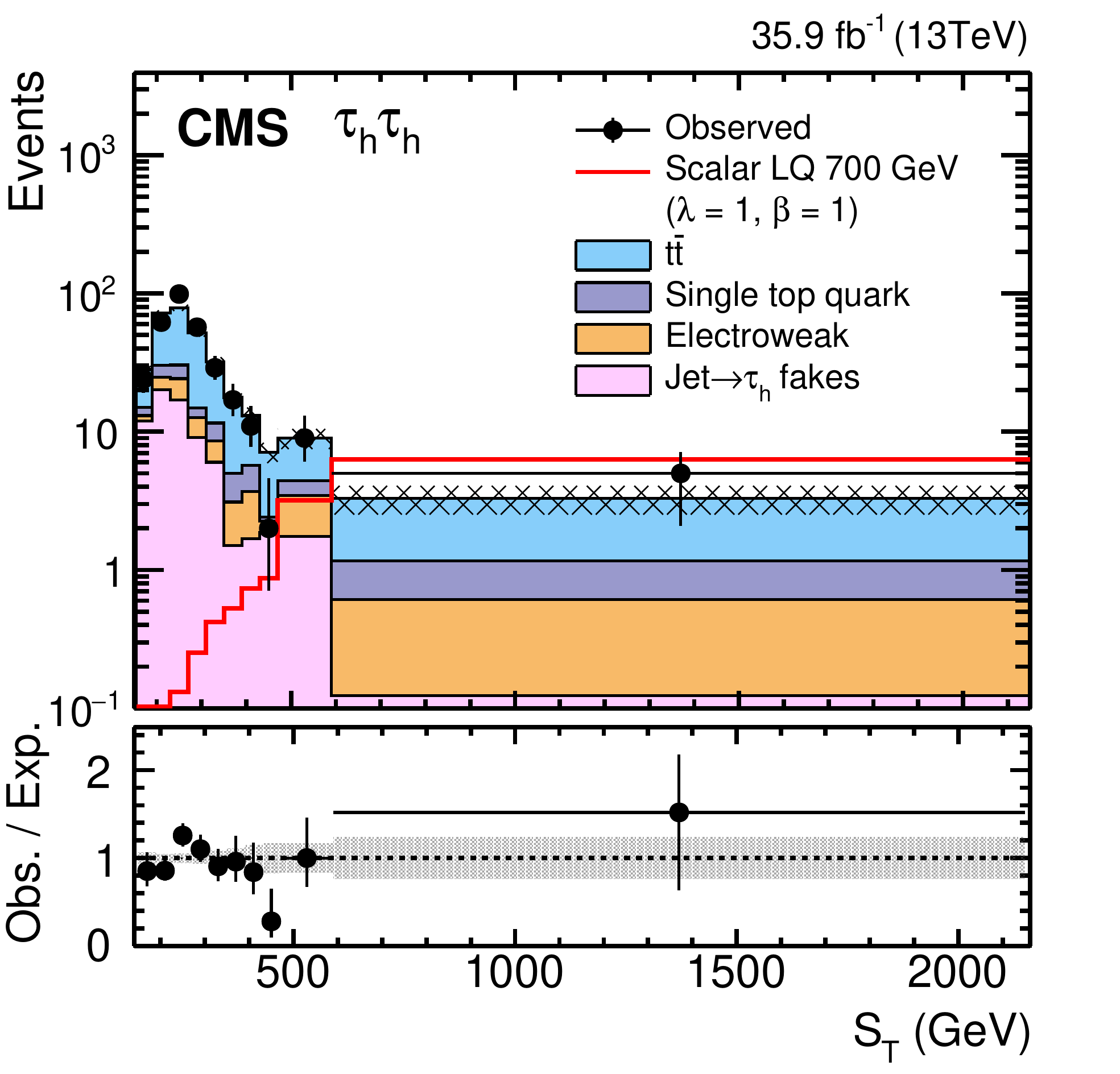}
     \includegraphics[width=.502\textwidth]{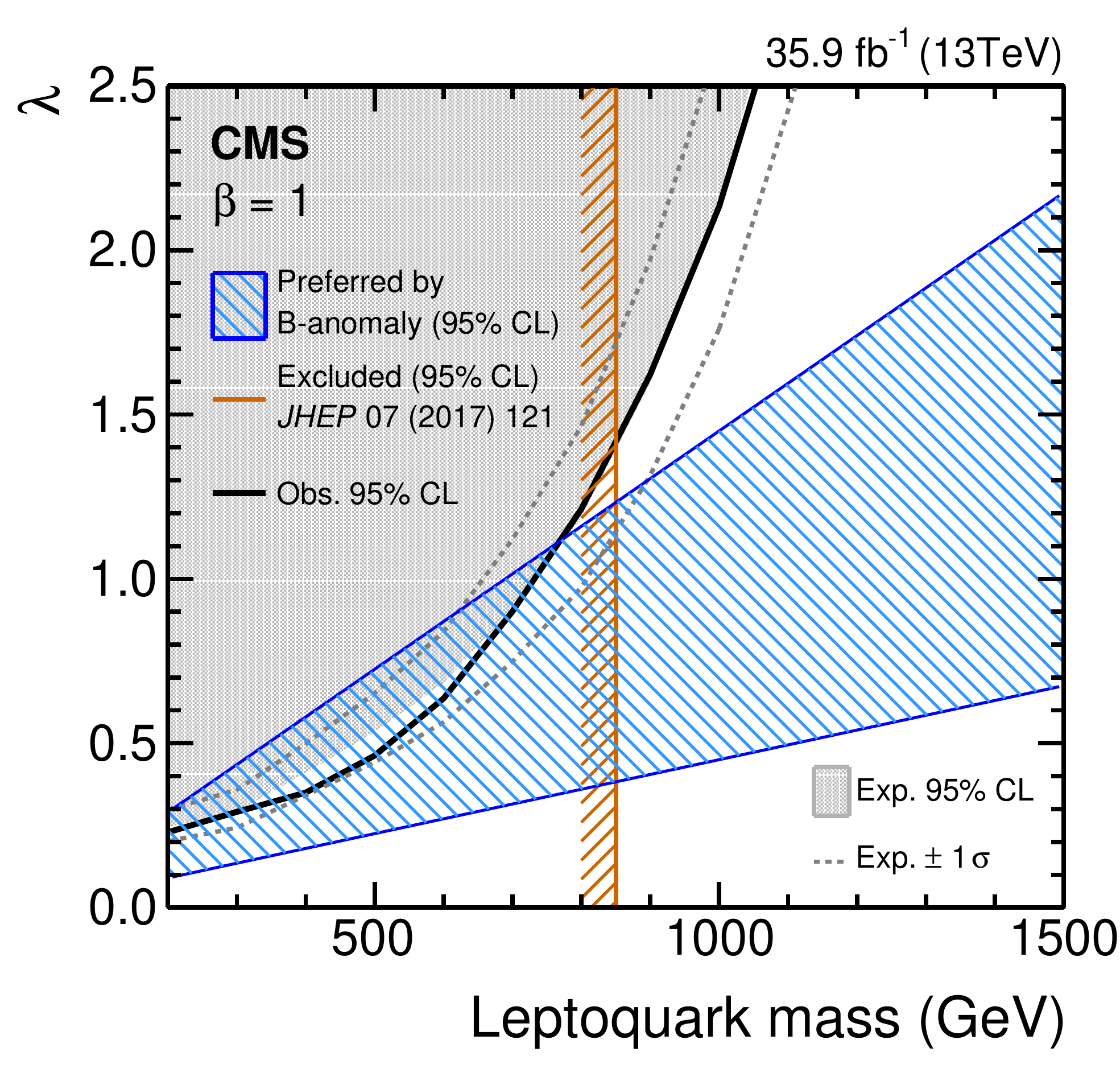}
     \caption{Search for singly produced LQs with
       couplings to $\tau$ leptons and bottom
       quarks~\cite{Sirunyan:2018jdk}: (left) observed $S_{\rm T}$
       distribution in the $\tau_h \tau_h $ signal region compared to
       the expectation from SM background processes and example signal
       contributions assuming a unit Yukawa coupling $\lambda=1$ at
       the LQ-lepton-quark vertex and a unit branching fraction
       $\beta=1$ of the LQ to a $\tau$ lepton and a bottom quark;
       (right) exclusion limits at 95\% CL on the
       Yukawa coupling $\lambda$ as a function of the LQ mass assuming
       $\beta=1$.}
     \label{fig:singleLQ}
\end{figure}

For LQs with couplings to third generation quarks, as considered in
this article, the single LQ production mode via gluon-quark scattering
is suppressed as it requires a bottom or top quark in the initial
state. For this reason, the single-production cross-section for LQs
with couplings to top quarks is negligibly small at the LHC. However,
since the cross-section depends on the Yukawa coupling $\lambda$ at
the LQ-lepton-quark vertex, it could be sizable for high values of
$\lambda$ for LQs with couplings to bottom quarks. The CMS
collaboration has performed a search for singly produced LQs decaying
to a $\tau$ lepton and a bottom quark by studying the associated
production of the LQ and a $\tau$ lepton~\cite{Sirunyan:2018jdk} and
compared the results with theory predictions designed to provide an
explanation for the observed anomalies in the
$B$-sector~\cite{Buttazzo:2017ixm}.

The analysis makes use of events with two $\tau$ candidates and at
least one $b$-tagged hadronic jet. Three categories are studied:
$\tau_{\rm h}\tau_{\rm h}$, $e \tau_{\rm h}$, and $\mu \tau_{\rm h}$,
among which $\tau_{\rm h}\tau_{\rm h}$ provides the highest
sensitivity. An additional $e\mu$ category is used to constrain the
dominant $t\overline{t}$ background contribution in a binned ML fit of
the $S_{\rm T}$ distributions which is performed as statistical
test. An example $S_{\rm T}$ distribution of the most sensitive
category is shown in Fig.~\ref{fig:singleLQ}\,(left). No significant
deviation from the SM expectation is observed in any of the categories
and exclusion limits on the cross-section as a function of $\lambda$,
$\beta$ and $M_{\rm LQ}$ can be determined and compared to the theory
prediction. The region excluded at 95\% CL for scalar LQs and
$\beta=1$ is shown in Fig.~\ref{fig:singleLQ}\,(right) in the ($M_{\rm
  LQ}$, $\lambda$)-plane. The analysis is able to exclude the region
of high values of $\lambda$ at small $M_{\rm LQ}$, while the CMS limit
from LQ pair-production in the $\tau b\,\tau b$
mode~\cite{Sirunyan:2017yrk} (see Sect.~\ref{sec:taub}) excludes the
region with $M_{\rm LQ}$<850\,GeV for all values of $\lambda$ as
indicated in the figure. In the meantime, the $\tau b\,\tau b$
pair-production limit has been improved by CMS to $M_{\rm
  LQ}<1020$\,GeV~\cite{CMS:2018eud}. The CMS analyses exclude a
significant part of the parameter space which has been identified as a
possible explanation for the $B$-anomalies~\cite{Buttazzo:2017ixm}
shown by the hatched area.

\section{Summary}

Since models of leptoquarks with couplings to the third generation of
SM quarks could provide a possible explanation of the anomalies
observed in the flavour sector, the CMS collaboration has performed a
broad range of different searches for the production of these states
using the LHC run-2 dataset. The data are in agreement with the SM
prediction. Certain parameter regions can be excluded which have been
identified as a possible explanation for the flavour
anomalies. Exclusion limits on masses of the leptoquark states in the
region of 1.0\,TeV and 1.8\,TeV have been obtained.  With future CMS
datasets expected for the high-luminosity phase of the LHC (HL-LHC,
$3000\,{\rm fb}^{-1}$), an increase of the accessible LQ mass range of
about 500\,GeV is expected.


\end{document}